\documentclass[twocolumn,preprintnumbers,amsmath,amssymb,prl]{revtex4}

\usepackage{xspace}
\usepackage{graphicx}% Include figure files
\usepackage{dcolumn}% Align table columns on decimal point
\usepackage{bm}% bold math
\usepackage{tabularx}% Table width
\usepackage{amsmath}% Higher math
\usepackage{color}%color

\begin{document}

\title{Evolution of the Nematic Susceptibility in LaFe$_{1-x}$Co$_x$AsO}
\author{Xiaochen Hong$^{1,*}$, Federico Caglieris$^1$, Rhea Kappenberger$^1$, Sabine Wurmehl$^1$, Saicharan Aswartham$^1$, Francesco Scaravaggi$^{1,2}$, Piotr Lepucki$^1$, Anja U. B. Wolter$^1$, Hans-Joachim Grafe$^1$, Bernd B\"{u}chner$^{1,2,3,4}$ and Christian Hess$^{1,3,\dag}$}

\affiliation{$^1$Leibniz-Institute for Solid State and Materials Research (IFW-Dresden), 01069 Dresden, Germany\\
$^2$Institute of Solid State and Materials Physics, TU Dresden, 01069 Dresden, Germany\\
$^3$Center for Transport and Devices, TU Dresden, 01069 Dresden, Germany\\
$^4$W\"{u}rzburg-Dresden Cluster of Excellence $ct.qmat$, TU Dresden, 01062 Dresden, Germany}

\date{\today}

\begin{abstract}
%%The identification of electronic nematicity across series of iron-based superconductors raises the question of its relationship with superconductivity and other ordered states. Here,
We report a systematic elastoresistivity study on LaFe$_{1-x}$Co$_x$AsO single crystals, which have well separated structural and magnetic transition lines. All crystals show a Curie-Weiss-like nematic susceptibility in the tetragonal phase. The extracted nematic temperature is monotonically suppressed upon cobalt doping, and changes sign around the optimal doping level, indicating a possible nematic quantum critical point beneath the superconducting dome. The amplitude of the nematic susceptibility shows a peculiar double-peak feature. This could be explained by a combined effect of different contributions to the nematic susceptibility, which are amplified at separated doping levels of LaFe$_{1-x}$Co$_x$AsO.
\end{abstract}

\pacs{not needed}

\maketitle
Unconventional superconductivity is intimately related to other electronic symmetry breaking states.
A common feature among unconventional superconductors is that by varying a tuning parameter $x$, such as pressure or chemical substitution, the superconductivity exists under a dome-like region in the $T-x$ phase diagram, and the extrapolation of some transition line hits zero temperature inside or close to that dome \cite{Pfleiderer,Keimer,Stewart,DaiPC}.
The formation of a spin density wave is the most recognized order in the normal state of unconventional superconductors, and its ubiquitousness stimulated theories to explain the origin of unconventional superconductivity mediated by fluctuations of the magnetic order \cite{Scalapino,Louis,Lohneysen,Pines,Schmalian,Sachdev}.
Recently, the report of newly recognized electronic orders beside magnetism in unconventional superconductors is infectious \cite{Keimer,Fradkin}, including nematicity, an electronic ordered state that spontaneously breaks the rotational symmetry of its host crystal \cite{Kivelson,Fradkin,Lilly}.
Nematic fluctuation by themselves are considered as possible mediator of electronic pairing \cite{Lederer,Metlitski,Maier,Labat,Maslov}.
Indeed, a nickel pnictide Ba$_{1-x}$Sr$_x$Ni$_2$As$_2$ was just discovered to be a nematicity-boosted superconductor \cite{JP}.

In the case of iron-based superconductors, the stoichiometric parent compounds are always antiferromagnetically (AFM) ordered below $T_N$, following a structural transition at $T_S$ \cite{Stewart,DaiPC}. The paramagnetic orthorhombic phase between $T_S$ and $T_N$ acquires highly anisotropic electronic properties regardless of the small lattice anisotropy \cite{FisherRev}.
In this electronic nematic phase, all ordered states are highly intertwined, making it hard to discern the leading instability among them. One can find clues by studying their pertinent fluctuations at higher temperatures \cite{Cano,FernendesSUST,NematicRev}. For nematicity, its fluctuations have been probed by various approaches, including the elastoresistivity measurements \cite{Chu2012,RSI}.
The divergent nematic susceptibility has been found in different families of iron-based compounds \cite{Kuo2016,Hosoi,LiSL}. These facts clearly eliminated the structural instability as the driving force for other transitions \cite{Chu2012,NematicRev}.

One prevailing understanding of the nematicity in iron-based compounds is to treat it as the result of fluctuating AFM order \cite{CFang,FeSe,Hackl,Dai}. This scenario received strong support from the discovery of a good scaling between the lattice softening and the magnetic fluctuating amplitude in Ba(Fe$_{1-x}$Co$_x$)$_2$As$_2$ over a wide doping range \cite{scaling}. As illustrated in Fig. 1(a), the $T_S$ and $T_N$ transition lines are very close to each other in Ba(Fe$_{1-x}$Co$_x$)$_2$As$_2$. The extension of these lines penetrate into a sign-reversed $s$-wave superconducting dome \cite{Stewart}, which is the fingerprint of AFM fluctuation driven superconductivity \cite{Inosov,LiS}.
In stark contrast, as Fig. 1(b) shows, there is no static magnetic order over the whole phase diagram of FeSe$_{1-x}$S$_x$. Theories proposed a different kind of superconductivity of sign-preserved pairing, based on unequal orbital occupancy \cite{Brink,Phillips,KuW,Devereaux,Onari}.
The nuclear magnetic resonance measurements saw no slowing down of spin fluctuations above $T_S$ in FeSe and much smaller changes of spin-lattice relaxation rates across $T_S$ than in other iron-based superconductors \cite{3pb,Baek}, lent support to the orbital scenarios and alternative origin of nematicity.
However, subsequent neutron scattering studies can identify spin fluctuations above $T_S$ \cite{QiSi}. Moreover, these spin fluctuations are found changing across $T_S$ and being anisotropic in the nematic phase \cite{QiSi2, Chen}. Thus, the importance of magnetic fluctuations is restored in FeSe. The leading instability and therefore the primary pairing mediator of the seemingly simple FeSe system remain elusive.

In this letter, we report the study of nematicity of a series of LaFe$_{1-x}$Co$_x$AsO single crystals. As Fig. 1(c) indicates, the $T_S$ and $T_N$ of LaFe$_{1-x}$Co$_x$AsO are well separated \cite{parent,Grafe,WangLR}. Superconductivity emerges after the total suppression of the AFM order \cite{Grafe,Francesco}, making LaFe$_{1-x}$Co$_x$AsO interesting to search for separated magnetic and nematic fluctuation regions.
By utilizing elastoresistivity measurements, the persistence of a Curie-Weiss-like nematic susceptibility deep in the tetragonal phase of LaFe$_{1-x}$Co$_x$AsO was found.
The Curie-Weiss temperature changes sign around the optimal doping level $x$ = 0.06, with an enhanced nematic susceptibility on its top. This indicates the existence of a nematic quantum critical point (nQCP) under the superconducting dome, and supports the proposals of nematicity as the driver for superconductivity \cite{Lederer,Metlitski,Maier,Labat,Maslov}.
Besides, another enhancement of the nematic susceptibility is resolved around $x$ = 0.04, close to the end point of the AFM order.

\begin{figure}
\includegraphics[clip,width=0.48\textwidth]{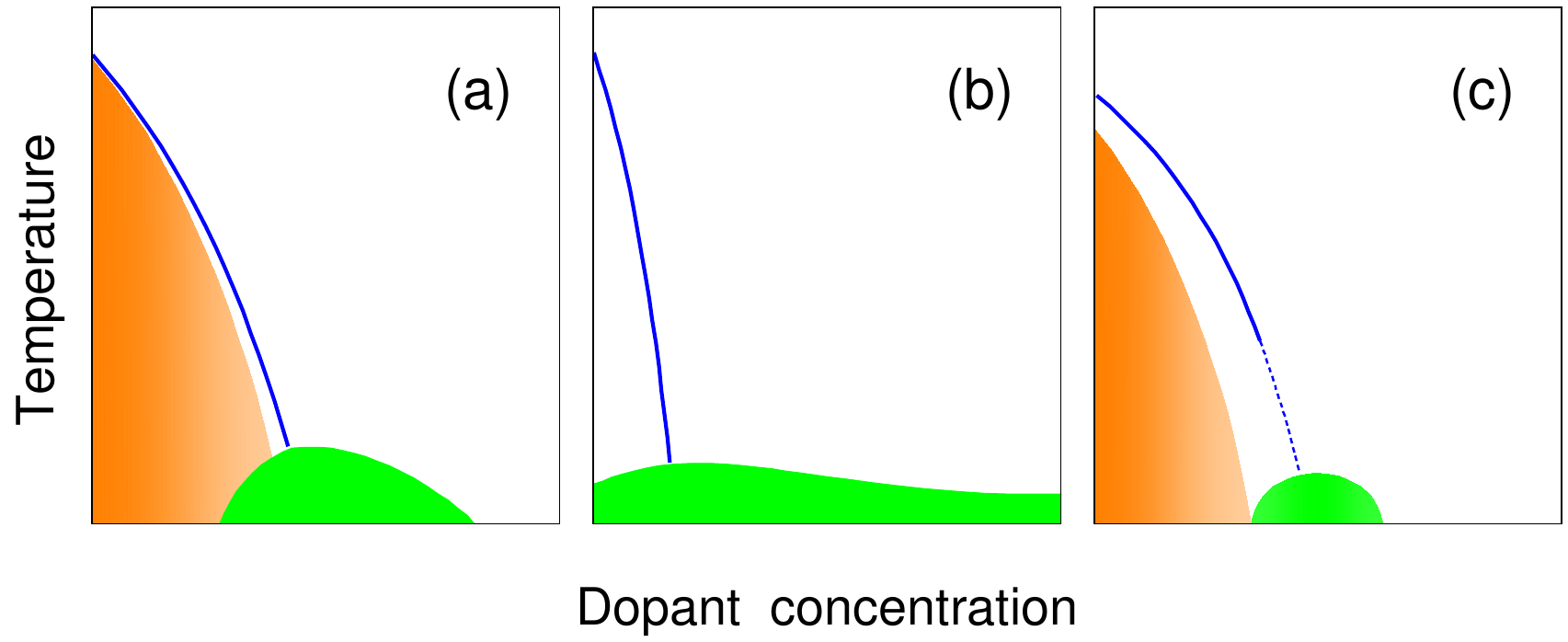}
\caption{(Color online).
Schematic $T-x$  phase diagrams of three representative iron based superconductors \cite{Stewart,Hosoi,Grafe,Francesco}: (a) Ba(Fe$_{1-x}$Co$_x$)$_2$As$_2$, (b) FeSe$_{1-x}$S$_x$, (c) LaFe$_{1-x}$Co$_x$AsO. The orange area is the AFM ordered state, the green dome is the superconducting state, and the solid blue line shows the structural transition (with dashed line as a guide to the eye). The nematic order exists below the structural transition line and above the AFM region. %anti-ferromagnetically
}
\end{figure}

\begin{figure}
\includegraphics[clip,width=0.48\textwidth]{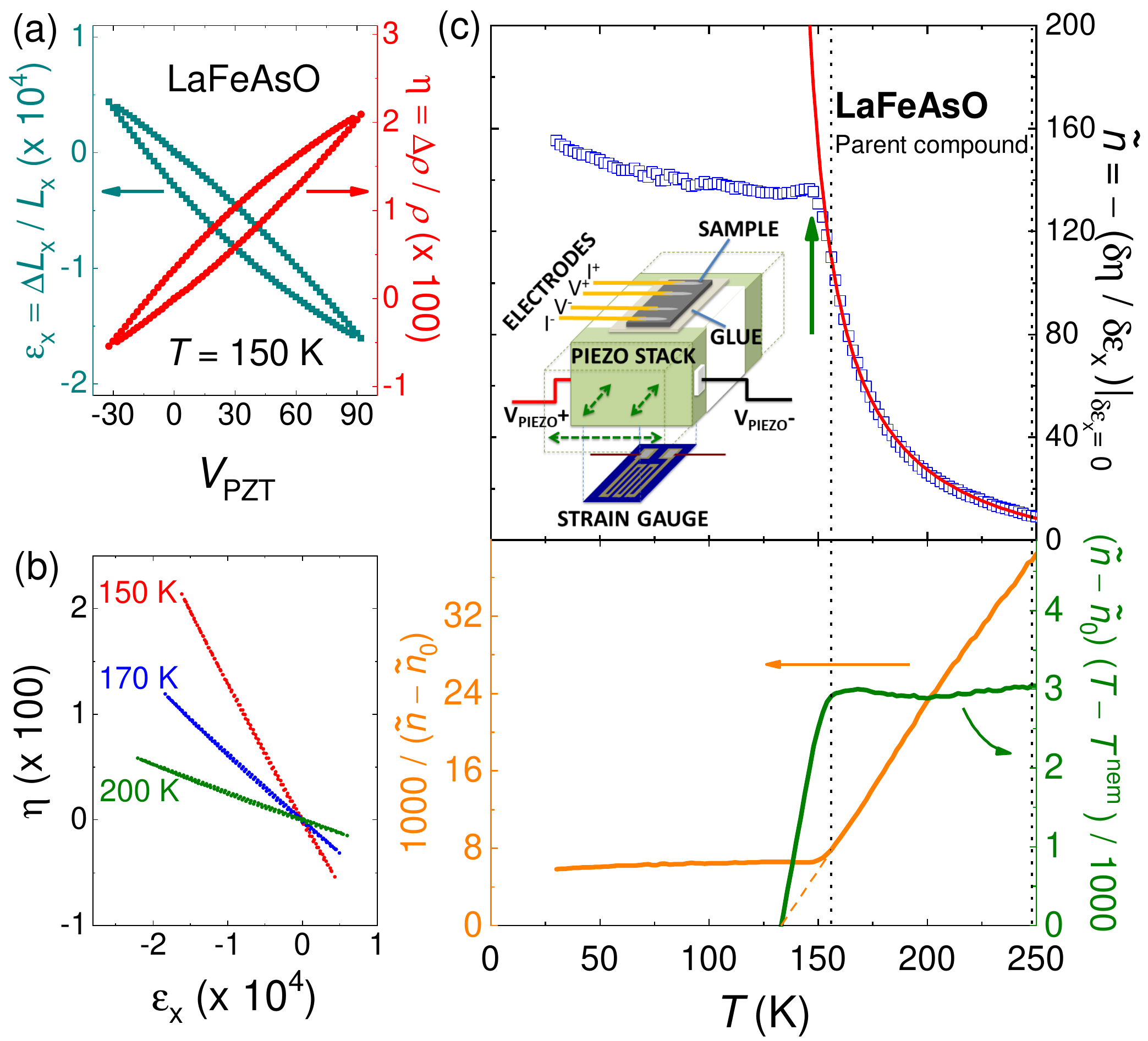}
\caption{(Color online). Elastoresistance measurements of the parent compound LaFeAsO.
(a) The relative change of strain ($\varepsilon_x = \Delta L_x / L_x$) and sample resistivity ($\eta = \Delta\rho / \rho$) according to the voltage applied to the piezo actuator at a fixed temperature $T = $ 150 K. The strain was measured along the current direction, which was aligned against the [110] axis of the crystal.
(b) The relationship between $\varepsilon_x$ and $\eta$ at some representative temperatures. The nematic susceptibility was obtained from $\tilde{n} = - (\delta\eta / \delta\varepsilon_x)$ in the $\varepsilon_x = 0$ limit.
(c) The temperature dependence of $\tilde{n}$ is shown as blue open squares in the upper panel. The structural transition temperature determined in Ref. \cite{parent} is indicated by the green arrow. The red curve represents a Curie-Weiss fitting to data in the temperature range between the vertical dotted lines. Inset is a schematic of the experimental setup. The inverse nematic susceptibility $(\tilde{n} - \tilde{n}_0)^{-1}$ and the Curie constant $C = (\tilde{n} - \tilde{n}_0)(T - T^{nem})$ are shown in the lower panel, indicating the fitting quality.}
\end{figure}

\begin{figure*}
\includegraphics[clip,width=0.99\textwidth]{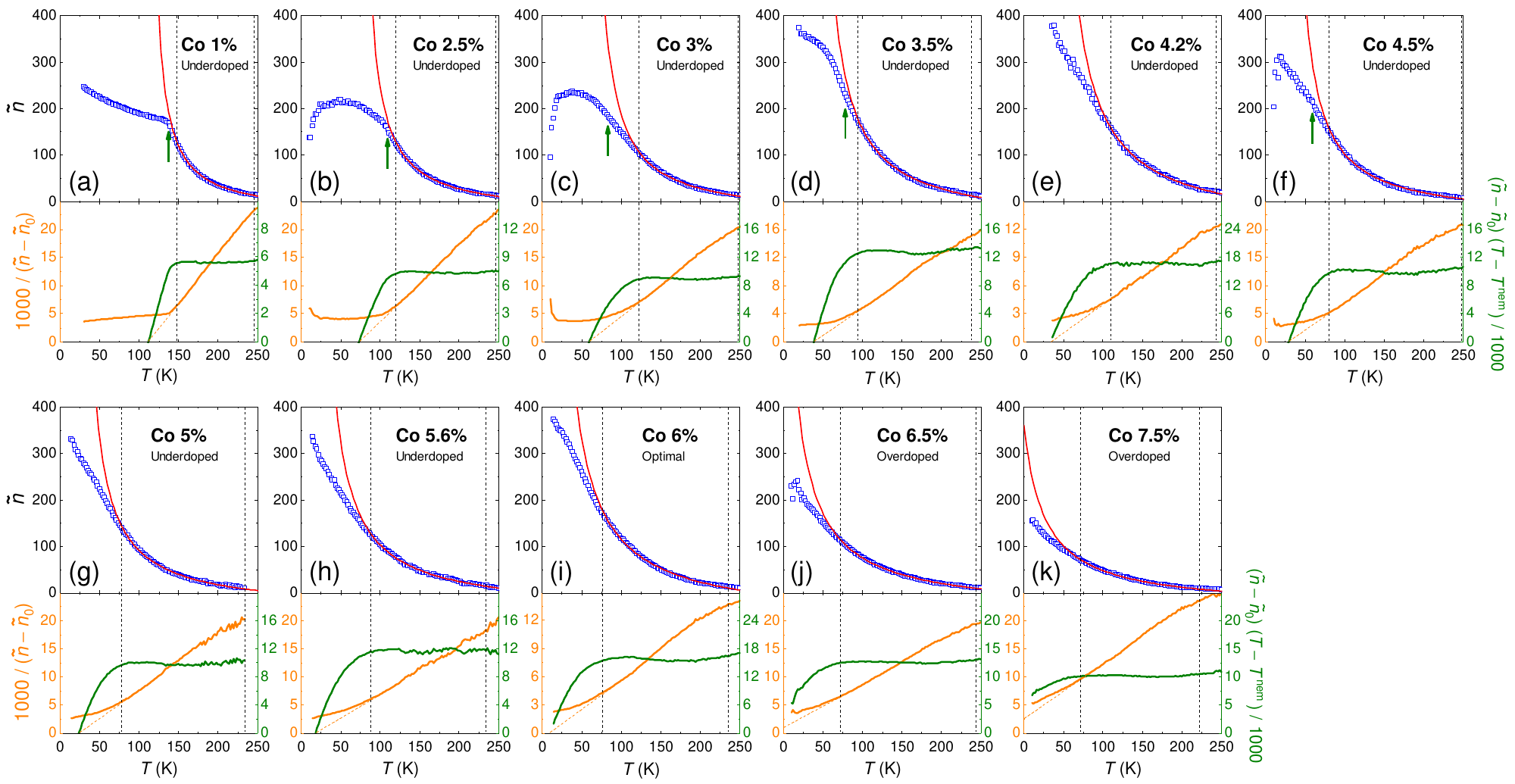}
\caption{(Color online). Divergent nematic susceptibility in LaFe$_{1-x}$Co$_x$AsO single crystals. Similar data as Fig. 2(c) for eleven samples of different doping level ranging from underdoped to overdoped is shown in (a - k). Cobalt content $x$ is the nominal composition, which agrees well with the actual concentration determined by energy dispersive X-ray spectroscopy on crystals from the same batches. The temperatures between which Curie-Weiss fitting was performed were determined by minimizing the systematic deviation of the Curie constant (see Ref. \cite{Kuo2016}). The structural transition temperature is indicated by the green arrow and is extracted from the references \cite{Francesco,WangLR}. There is no arrow in Fig. 3(e) because the Co 4.2\% sample was not studied in the references \cite{Francesco,WangLR}.}
\end{figure*}

LaFe$_{1-x}$Co$_x$AsO single crystals were prepared by the solid state single crystal growth method \cite{parent}. The oriented samples were cut into a typical size of 1.2 mm $\times$ 0.5 mm in plane and cleaved to around 20  $\mu$m thickness to ensure an efficient strain transmission to the sample \cite{Chu2012}. The experimental setup is illustrated in the insert of Fig. 2(c). The sample was glued to a commercial piezoelectric actuator (Piezomechanik PSt 150 / 5$\times$5$\times$7) by using an adhesive epoxy (Devcon, No. 14250), after electrically contacting voltage and current leads directly on the fresh surface with sliver paint. A strain gauge was glued on the other side of the actuator and was measured with a built-in Wheatstone bridge circuit. To eliminate a possible temperature change caused by driving the piezoelectric actuator, data were taken after waiting several seconds at each strain step.

Elastoresistivity measured along the [110] direction of the LaFeAsO single crystal is shown in Fig. 2. By changing the voltage across a piezo actuator, one can precisely tune the strain applied to the sample glued on the actuator's edge.
In the nematic ordered state, the resistivity anisotropy of the in-plane axes is very large \cite{Chu2010,Tanatar}. This anisotropy represents the electronic nematicity. Due to the electron-lattice coupling, strain can induce a resistivity anisotropy above the nematic ordered state if nematic fluctuations exist \cite{Chu2012}.
As illustrated in Fig. 2 (b), the fractional change of sample resistivity ($\eta = \Delta\rho / \rho$) shows a perfect linear relationship to the strain ($\varepsilon_x = \Delta L_x / L_x$) applied along the current direction. As demonstrated in the references \cite{Chu2012,RSI,Kuo2016,Hosoi}, the slope of $\eta(\varepsilon_x)$ curve probes the nematic susceptibility of the sample. The negative slope means that the in-plane resistivity of LaFeAsO is higher along the shorter orthorhombic direction. For simplicity, we use $- (\delta\eta / \delta\varepsilon_x)$ in the small strain limit as the definition of $\tilde{n}$, which is a gauge of the nematic susceptibility.
The temperature dependence of $\tilde{n}$ in LaFeAsO is shown in Fig. 2(c). A clear kink is resolved at the structural transition $T_S$, above which the $\tilde{n}(T)$ curve can be well fitted to the Curie-Weiss type temperature dependence
\begin{equation}
\tilde{n} = \tilde{n}_{0} + \frac{C}{T - T^{nem}},
\end{equation}
in which $\tilde{n}_{0}$ is the intrinsic piezoresistivity effect unrelated to electronic nematicity, $T^{nem}$ is the nematic transition temperature in the mean field theory, and $C$ is the Curie constant which indicates the magnitude of the nematic susceptibility.

Similar measurements and analyses were conducted for LaFe$_{1-x}$Co$_x$AsO single crystals across the substitution series ($x$ up to 0.075). As shown in Fig. 3, for all the samples we have studied, their $\tilde{n}(T)$ curves diverge in a clear Curie-Weiss-like form. The kink at $T_S$ becomes less pronounced with doping and is unresolvable in our data set for $x > 0.03$, while $T_S$ could be traced up to $x = 0.045$ in thermal expansion measurements (indicated by green arrows) \cite{Francesco}.
%%and up to $x = 0.056$ in a nuclear magnetic resonance study \cite{Grafe}.
One may attribute the smearing of the $T_S$ feature to the increasing impurity scattering effect. In such a case, according to Ref. \cite{Kuo2014}, $\tilde{n}$ should remain practically unaffected by the defects and impurity concentrations in the tetragonal phase.
However, the blurred feature of $T_S$ could also be an intrinsic property of LaFe$_{1-x}$Co$_x$AsO, since the structural transition anomaly observed in the thermal expansion experiments becomes broader with increasing $x$ \cite{Francesco}.
Indeed, a nuclear quadrupole resonance study on LaFe$_{1-x}$Co$_x$AsO showed that competing local charge environments exist at the nanoscale in the underdoped region \cite{Grafe}, similar to the sister compound LaFeAsO$_{1-x}$F$_x$ \cite{Lang,Lang2}. In that sense, it is natural that the features of $T_S$ from averaged properties are muddled by doping.

We analyze the $\tilde{n}(T)$ data for the finite doping levels in the same Curie-Weiss manner as for the parent compound (solid red lines).
In order to choose the proper fitting range of the $\tilde{n}(T)$ curves, especially for those without (a clear) $T_S$, we followed the procedure described in Ref. [\onlinecite{Kuo2016}]. Additionally, the fitting range is allowed to alter by 20 K for each sample in order to estimate the uncertainty caused by improper fitting ranges \cite{SM}, set as error bars of the fitted parameters $T^{nem}$ and $C$ in the following discussion.

\begin{figure}
\includegraphics[clip,width=0.49\textwidth]{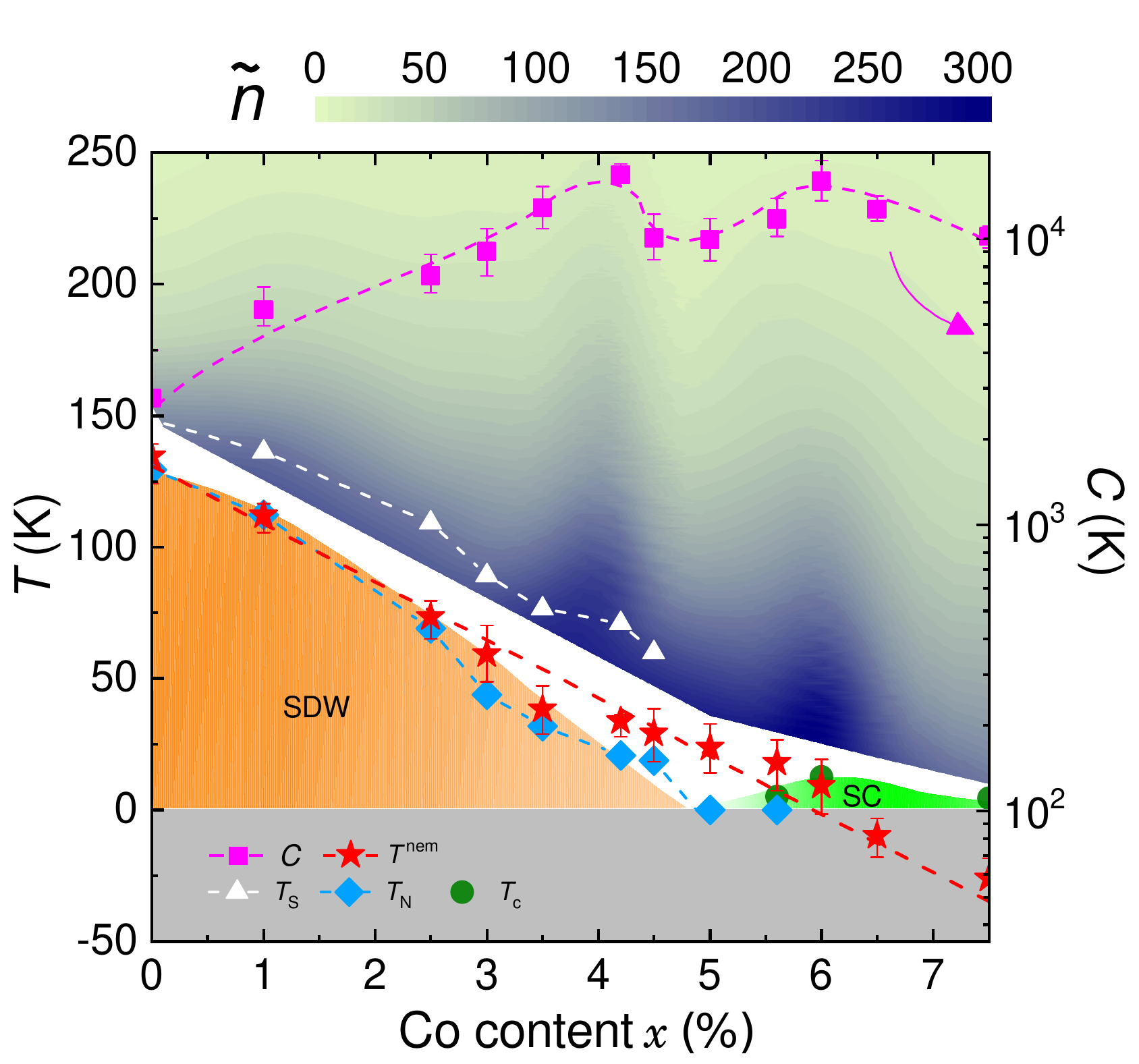}
\caption{(Color online). Phase diagram and nematic susceptibility of LaFe$_{1-x}$Co$_x$AsO.
The white triangles, light blue diamonds, and green circles indicate $T_S$, $T_N$, and $T_c$ respectively, according to the references \cite{Francesco,Grafe,WangLR}. The orange area encloses the AFM ordered state, while the superconducting region is represented by the green area.
The nematic critical temperature $T^{nem}$ (red star) and the divergency coefficient (Curie constant) $C$ (magenta square, right axis, in log scale) are obtained by the Curie-Weiss fits of the nematic susceptibility data shown in Fig. 2 and Fig. 3.
The error bars of them come from altering the fit range (see text). The dashed lines are guides to the eye.
Evolution of the magnitude of $\tilde{n}$ is shown by the color plot, which shows a double-peak feature.}
\end{figure}

The information extracted from Fig. 3 is summarized in Fig. 4, plotted together with the phase diagram of LaFe$_{1-x}$Co$_x$AsO derived from the doping dependence of $T_S$, $T_N$, and $T_c$ \cite{Grafe,Francesco}.
First of all, one can see that the nematic transition temperature $T^{nem}$ is linearly suppressed by doping, and finally changes its sign around the optimal doping level $x \approx 0.06$.
It is known that superconductivity could get boosted via fluctuations related to electronic instabilities.
The most studied case is that of AFM fluctuations with certain wave vectors around a magnetic critical point.
As shown in Fig. 1(c), the magnetic transition is totally suppressed before the superconductivity emerges in LaFe$_{1-x}$Co$_x$AsO \cite{Grafe,Francesco}, suggesting that the magnetic fluctuations near the critical point may not contribute to the superconducting transition temperature $T_c$ directly.
Theoretically, a nQCP and its pertinent fluctuation can also promote the superconductivity via the attractive forces in all paring channels \cite{Lederer,Metlitski,Maier,Labat}, or by reducing the intra-pocket repulsion \cite{FernendesSUST,Maslov}.
Our observation of the sign change of $T^{nem}$ around $x$ = 0.06 strongly corroborates those theoretical expectations.

Interestingly, upon investigating the doping dependence of the magnitude of $\tilde{n}$, one can clearly observe a significant enhancement of this quantity at the optimal doping level.
This can be inferred from the data in the upper panels of Fig. 3(a-k) and also the contour plot presented in Fig. 4, revealing that the enhanced $\tilde{n}$ mimics the superconducting dome.
The latter trend is also clearly visible in the evolution of the Curie constant displayed in Fig. 4, too, which represents the divergency coefficient of the $\tilde{n}(T)$ curve.
This finding of an enhanced $\tilde{n}$ on top of the critical point of $T^{nem}$ and maximum $T_c$ strongly underpins the above notion of a putative nQCP near the optimal doping level.

The further inspection of $\tilde{n}(T)$ in Figs. 3 and 4 at lower doping clearly reveals a second peak-like enhancement of it, very close to the magnetic critical point where the AFM order disappears but still at doping levels with an AFM ground state.
An alternative nematic critical point as the origin of the additional peak in the underdoped region at $x$ $\approx$ 0.04 can be ruled out because the fitting of $\tilde{n} (T)$ gives unambiguously a finite $T^{nem} \approx$ 35 K at $x$ $\approx$ 0.04 \cite{SM}.
One may speculate that the enhanced $\tilde{n}$ around $x$ = 0.04 is caused by magnetic fluctuations, which are critically enhanced at the doping levels just below the disappearance of the AFM order (like in many other iron-based  superconductors \cite{DaiPC}), and remain small up to the optimal doping level $x$ = 0.06 \cite{Grafe}.
This notion is supported by the established $T^{nem} \approx$ 35 K for $x \approx$ 0.04 which roughly matches the observed magnetic ordering temperature $T_N$ around this doping level.
Thus, altogether our data seem to reveal that in the underdoped region of the phase diagram for $x \leq$ 0.05, $\tilde{n}$ follows the trend of the magnetic fluctuations which are connected to the incipient order.
At higher doping, $\tilde{n}$ is enhanced and seems decoupled from the magnetic fluctuations, while on top of the presumptive nQCP, $\tilde{n}$ and $T_c$ domes exist despite a reduced magnetic susceptibility.
It is worthwhile to point out that the seeming appearance of two separated criticalities suggests nematicity itself can exist as a primary fluctuation and may be responsible for the superconductivity.
Furthermore, the comparable Curie constant $C$ around the two critical doping levels indicates a similar strength of the elasto-electronic and elasto-magnetic couplings in this material. This is an interesting point to look at in further investigations.

The dichotomic-origin of the $\tilde{n}$ in LaFe$_{1-x}$Co$_x$AsO provides fresh input for rationalizing the importance of nematic fluctuations in other members of the iron-based superconductor family.
In electron-doped BaFe$_2$As$_2$, the peak of $\tilde{n}$ has a broad appearance with the tendency to maximize at a slightly underdoped level \cite{Chu2012,Kuo2016,Christoph}.
Since the magnetic and structural transition lines of those materials are close to each other, the measured $\tilde{n}R$, a superposition of the two profiles which maximize at slightly separated end points, could appear like a broadened peak in between.
We note that FeSe$_{1-x}$S$_x$ has recently been proven to be a remarkable system which possesses an nQCP under its superconducting region \cite {Hosoi,Licciardello}. Although no AFM order exists in its ambient pressure phase diagram, magnetic fluctuations are still interpreted as the booster of superconductivity in FeSe$_{1-x}$S$_x$ \cite{Matsuura}.
As we have mentioned above, however, magnetic fluctuations are suppressed at the superconducting doping levels in LaFe$_{1-x}$Co$_x$AsO, suggesting it is a more suitable prototype system to search for the relationship between superconductivity and nematicity.
Finally, we point out that recent reports also claimed separated QCPs masked below the wide-spread superconducting dome in cobalt- and nickel-doped NaFeAs, extracted from the nuclear magnetic resonance and neutron scattering spectra respectively \cite{Zheng, Wang}. We therefore suspect a similar double-peak $\tilde{n}$ profile would also be detected in the NaFeAs family.

To summarize, we found that the Curie-Weiss-like $\tilde{n}$ also exists in LaFe$_{1-x}$Co$_x$AsO, in line with the other iron-based superconductors.
The sign-change of $T^{nem}$ around $x$ = 0.06 and the divergent amplitude of $\tilde{n}$ by approaching this doping level are consistent with a possible nQCP at the optimal doping level.
An additional peak of $\tilde{n}$ in the underdoped region around $x$ $\approx$ 0.04 is present close to the end point of the AFM transition line in LaFe$_{1-x}$Co$_x$AsO, which we attribute to a coupling of the nematic fluctuation to critical magnetic fluctuations.
Altogether, the double-peak feature of $\tilde{n}$ suggests another origin of nematicity in iron-based compounds, besides the well accepted vestigial magnetism explanation \cite{CFang,FeSe}. A detailed map out of the magnetic fluctuations in LaFe$_{1-x}$Co$_x$AsO is important to further understand nematicity in this material.

We would like to thank R\"{u}diger Klingeler, Sven Sauerland, Liran Wang, Seung-Ho Baek, Christoph Wuttke, and Steffen Sykora for helpful discussions. This work has been supported by the Deutsche Forschungsgemeinschaft (DFG) through the SFB 1143 (project-id 247310070), through the Priority Programme SPP1458 (Grant  No.BU887/15-1), under grant DFG-GRK 1621, through DFG Research Grants AS 523/4-1 (S.A.) CA1931/1-1 (F.C.), and through the Emmy Noether Programme WU595/3-3 (S.W.). X.C.H. acknowledges the support of the Leibniz-DAAD Research Fellowship. This project has received funding from the European Research Council (ERC) under the European Union's Horizon 2020 research and innovation programme (grant agreement No. 647276-MARS-ERC-2014-CoG) and the W\"{u}rzburg-Dresden Cluster of Excellence on Complexity and Topology in Quantum Matter $ct.qmat$ (EXC 2147, project-id 390858490).

$^*$ E-mail: x.c.hong@ifw-dresden.de

$^\dag$ E-mail: c.hess@ifw-dresden.de

\end{document}